\pdfoutput=1

\documentclass[11pt,twoside,a4paper,cmspaper,final,collab]{cms-tdr}
\def\svnVersion{289262}\def\cmsCernNo{2013-037}\def\cmsCernDate{\today}\def\cmsMessage{Published in Physics Letters B as \href{http://dx.doi.org/10.1016/j.physletb.2015.04.060}{\doi{10.1016/j.physletb.2015.04.060.}}}

\begin{document}\cmsNoteHeader{TOP-13-010}

\hyphenation{had-ron-i-za-tion}
\hyphenation{cal-or-i-me-ter}
\hyphenation{de-vices}
\RCS$Revision: 282914 $
\RCS$HeadURL: svn+ssh://svn.cern.ch/reps/tdr2/papers/TOP-13-010/trunk/TOP-13-010.tex $
\RCS$Id: TOP-13-010.tex 282914 2015-03-31 17:12:53Z alverson $
\newlength\cmsFigWidth
\ifthenelse{\boolean{cms@external}}{\setlength\cmsFigWidth{0.85\columnwidth}}{\setlength\cmsFigWidth{0.4\textwidth}}
\ifthenelse{\boolean{cms@external}}{\providecommand{\cmsLeft}{top}}{\providecommand{\cmsLeft}{left}}
\ifthenelse{\boolean{cms@external}}{\providecommand{\cmsRight}{bottom}}{\providecommand{\cmsRight}{right}}
\ifthenelse{\boolean{cms@external}}{\providecommand{\xTolerance}{9999}}{\providecommand{\xTolerance}{600}}

\newcommand{\mue}{\ensuremath{\mu^\pm\cmsSymbolFace{e^\mp}}\xspace}
\newcommand{\emu}{\ensuremath{\cmsSymbolFace{e^\pm}\mu^\mp}\xspace}
\newcommand{\ttbb}{\ensuremath{\cPqt\cPaqt\cPqb\cPaqb}\xspace}
\newcommand{\ttcc}{\ensuremath{\cPqt\cPaqt\cPqc\cPaqc}\xspace}
\newcommand{\ttb}{\ensuremath{\cPqt\cPaqt\cPqb\cmsSymbolFace{j}}\xspace}
\newcommand{\ttll}{\ensuremath{\cPqt\cPaqt\cmsSymbolFace{LF}}\xspace}
\newcommand{\ttjj}{\ensuremath{\cPqt\cPaqt\cmsSymbolFace{jj}}\xspace}
\newcommand{\ttH}{\ensuremath{\cPqt\cPaqt\cmsSymbolFace{H}}\xspace}

\newcommand{\sttbb}{\ensuremath{\sigma_{\ttbb}}\xspace}
\newcommand{\sttjj}{\ensuremath{\sigma_{\ttjj}}\xspace}

\cmsNoteHeader{TOP-13-010}
\title{Measurement of the cross section ratio $\sttbb/\sttjj$ in $\Pp\Pp$ collisions at $\sqrt{s} = 8\TeV$}
\author{The CMS Collaboration}

\date{\today}

\abstract{
The first measurement of the cross section ratio $\sttbb/\sttjj$
is presented using a data sample corresponding to an integrated luminosity of 19.6\fbinv collected in pp collisions at $\sqrt{s} = 8\TeV$
with the CMS detector at the LHC.
Events with two leptons ($\Pe$ or $\Pgm$) and four reconstructed jets, including two identified as b quark jets, in the final state are selected.
The ratio is determined for a minimum jet transverse momentum \pt of both 20 and 40\GeVc.
The measured ratio is $0.022 \pm 0.003\stat\pm0.005\syst$ for $\pt > 20\GeVc$.
The absolute cross sections $\sttbb$ and $\sttjj$ are also measured.
The measured ratio for $\pt> 40\GeVc$ is compatible with
a theoretical quantum chromodynamics calculation at next-to-leading order.
}

\hypersetup{%
pdfauthor={CMS Collaboration},%
pdftitle={Measurement of the cross section ratio sigma(t t-bar b b-bar) / sigma(t t-bar jj) in pp collisions at sqrt(s) = 8 TeV},%
pdfsubject={CMS},%
pdfkeywords={CMS, physics, top physics}}

\maketitle

\section{Introduction}\label{sec:intro}
With the observation of a new boson at a mass around 125\GeVcc~\cite{atlas:newhiggs,cms:newhiggs,cms:newhiggsdetail}
whose properties are consistent with those of
the standard model (SM) Higgs boson \PH~\cite{cms:propertieshiggs_fourlepton,cms:higgstoWW,cms:Htautau,cms:VHbb,atlas:newhiggscoupling,atlas:newhiggsevidence},
the SM appears to be complete.
One of the most sensitive channels in the discovery of the Higgs boson, \HGG, is expected to have top quark loops both in the production and decay of the Higgs boson in the SM.
Hence, it is important to determine the couplings of the new boson to fermions, especially to the top quark.
In the SM, one of the most promising channels for a direct measurement of the top quark Yukawa coupling
is the production of the Higgs boson in association with a \ttbar pair (\ttH), where
the Higgs boson decays to \bbbar, thus  leading to a \ttbb final state.

{\tolerance=\xTolerance
The expected quantum chromodynamics (QCD) cross section for \ttH production in pp collisions at $\sqrt{s}= 8\TeV$,
calculated to next-to-leading order (NLO),
is $0.128^{+0.005}_{-0.012}\,\text{(scale)}\pm 0.010\unit{pb} \,\text{(PDF+$\alpha_\mathrm{S}$)}$ \cite{yellowreport},
where the uncertainty labelled ``scale'' refers to the uncertainty from the factorization and renormalization scales ($\mu_F$ and $\mu_R$),
and the uncertainty labelled ``PDF+$\alpha_\mathrm{S}$'' comes from the uncertainties in the
parton distribution functions (PDFs) and the strong coupling constant $\alpha_\mathrm{S}$.
This final state, which has not yet been observed, has an irreducible
nonresonant background from the production of a top quark pair in
association with a b quark pair.
Calculations of the inclusive production cross section for \ttbar events with additional jets have been
performed to NLO precision~\cite{Bevilacqua:arXiv0907.4723,PhysRevD.84.114017,PhysRevLett.104.162002,Bredenstein:arXiv1001.4006,Worek:arXiv1112.4325,ttbbQCD}.
For a proton-proton centre-of-mass energy of 8\TeV, the predictions for the production of a top quark pair with two additional jets \ttjj and with two additional b quark jets \ttbb are
$\sttjj = 21.0 \pm 2.9\,\text{(scale)}\unit{pb}$ and $\sttbb = 0.23 \pm 0.05\,\text{(scale)}\unit{pb}$, respectively~\cite{ttbbQCD}.
In this calculation, the additional jets are required to have transverse momenta $\pt > 40\GeVc$ and absolute pseudorapidity $\abs{\eta}<2.5$,
while for the \ttH production value quoted above, no such requirements are applied to the decay products of the Higgs boson.
The dominant uncertainties in these calculations are from the factorization and renormalization scales~\cite{PhysRevLett.103.012002,JHEP:ttbb}
caused by the presence of two very different scales in this process, the top quark mass and the jet \pt.
Therefore, experimental measurements of $\sttjj$ and $\sttbb$ production can provide a good test of NLO QCD theory
and important input about the main background in the search for the \ttH process.
\par}

In this Letter, the first measurements of the cross sections \sttbb and \sttjj and their ratio are presented.
The analyzed data sample of pp collisions at a centre-of-mass energy of 8 \TeV was collected with the CMS experiment at the CERN LHC
and corresponds to an integrated luminosity of $19.6 \pm 0.5$\fbinv~\cite{CMS-PAS-LUM-13-001}.
The primary motivation for measuring the cross section ratio is that
many kinematic distributions are expected to be similar for \ttbb and \ttjj,
leading to reduced systematic uncertainties in the ratio.

\section{CMS detector and event reconstruction}\label{sec:det}

The central feature of the CMS apparatus is a superconducting solenoid of 6\unit{m} internal diameter, providing a magnetic field of 3.8\unit{T}. Within the superconducting solenoid volume are a silicon pixel and strip tracker, a lead tungstate crystal electromagnetic calorimeter (ECAL), and a brass and scintillator hadron calorimeter (HCAL), each composed of a barrel and two endcap sections. Muons are measured in gas-ionization detectors embedded in the steel flux-return yoke outside the solenoid. Extensive forward calorimetry complements the coverage provided by the barrel and endcap detectors.

The particle-flow event algorithm reconstructs and identifies each single particle with an optimized combination of all subdetector information~\cite{PAS_PFT-09-001,PAS_PFT-10-001}. The energy of photons is directly obtained from the ECAL measurement, corrected for zero-suppression effects. The energy of electrons is determined from a combination of the electron momentum at the primary interaction vertex as determined by the tracker, the energy of the corresponding ECAL cluster, and the energy sum of all bremsstrahlung photons spatially compatible with originating from the electron track. The energy of muons is obtained from the curvature of the corresponding track. The energy of charged hadrons is determined from a combination of their momentum measured in the tracker and the matching ECAL and HCAL energy deposits, corrected for zero-suppression effects and for the response function of the calorimeters to hadronic showers. Finally, the energy of neutral hadrons is obtained from the corresponding corrected ECAL and HCAL energy.

Jet momentum is determined as the vectorial sum of all particle momenta in the jet, and is found from simulation to be within 5 to 10\% of the true momentum over the whole \pt spectrum and detector acceptance.
An offset correction is applied to take into account the extra energy clustered in jets due to additional proton-proton interactions within the same bunch crossing (pileup).
Jet energy corrections are derived from simulation, and are confirmed with in situ measurements with the energy balance of dijet and photon+jet events. Additional selection criteria are applied to each event to remove spurious jet-like features originating from isolated noise patterns in certain HCAL regions.

A more detailed description of the CMS detector, together with a definition of the coordinate system used and the relevant kinematic variables, can be found in Ref.~\cite{Chatrchyan:2008zzk}.

\section{Simulation and definition of signal events}
Monte Carlo (MC) simulated data samples for the \ttbar signal are generated
by the \MADGRAPH~(v. 5.1.3.30) event generator~\cite{MADGRAPH}
with matrix elements (ME) at leading order, allowing up to three additional partons including b quarks.
The generated events are interfaced with \PYTHIA (v. 6.426)~\cite{pythia} to provide the showering of the partons, and
to perform the matching of the soft radiation with the contributions from the ME.
The $\tau$ lepton decays are handled with \TAUOLA~(v. 2.75)~\cite{tauola}. The \POWHEG~(v. 1.0) generator~\cite{Nason:2004rx,Frixione:2007vw,Alioli:2010xd}
at NLO, interfaced with \PYTHIA, is used for cross-checks and systematic studies.
A $\cPZ/\gamma^{*}$+jets background sample is simulated in \MADGRAPH.
The \ttH process is modelled using \PYTHIA.
The electroweak production of single top quarks ($\Pp\Pp \to \cPqt\PW$ and $\Pp\Pp\to\cPaqt\PW$) is simulated in \POWHEG
with an approximate next-to-next-to-leading-order (NNLO) cross section calculation~\cite{kidonakis:tw}.
The CTEQ6L1~\cite{Pumplin:2002vw} set of PDFs is used for the \MADGRAPH  and \PYTHIA samples,
while the CTEQ6M~\cite{PhysRevD.78.013004} set is used for the \POWHEG samples.
The CMS detector response is simulated using \GEANTfour~(v.~9.4)~\cite{Agostinelli:2002hh}.
The pileup distribution used in the simulation is weighted to match the one observed in data.

Measurements are reported for two different regions of the phase space: a visible phase space and the full phase space.
In the visible phase space, all $\ttbb$ final state particles ($\ttbb \to \cPqb\PWp\cPaqb\PWm\cPqb\cPaqb \to \cPqb{\ell}^{+} \Pgn \cPaqb{\ell}^{-} \Pagn \cPqb\cPaqb$) except the neutrinos,
\ie the charged leptons and jets originating from the decays of the top quarks,
as well as the two additional b quark jets (``b jets''), are required to be within the same experimentally accessible kinematic region.
Simulated \ttbb events are defined to be in the visible phase space and are categorized as coming from the \ttjj process
if they contain, at the generator level, at least four particle-level jets, including
at least two jets originating from b quarks, and two leptons ($\ttjj \to \cPqb\PWp\cPaqb\PWm\mathrm{jj}\to \cPqb{\ell}^{+} \Pgn\cPaqb {\ell}^{-} \Pagn \mathrm{jj}$).
Each lepton must have $\pt>20\GeVc$, $\abs{\eta}<2.4$, and come
from the decay of a W boson from one of the top quarks.
Electrons or muons originating from the leptonic decays of $\Pgt$ leptons produced in $\PW \to \Pgt\Pgn$ decays are included.
Jets which are within $\DR =\sqrt{\smash[b]{\Delta \phi^2+\Delta \eta^2}} < 0.5$ of an identified electron or muon are removed,
where $\Delta \phi$ and $\Delta \eta$ are the differences in azimuthal angle and pseudorapidity between
the directions of the jets and the lepton.
The particle-level jets are obtained by combining all final-state particles, excluding neutrinos, at the generator level with an anti-\kt clustering algorithm~\cite{antikt}
with a distance parameter of 0.5 and are required to satisfy $\abs{\eta} < 2.5$ and $\pt > 20$\GeVc, which is lower than the reconstructed minimum jet \pt,
as described below.
The b and c quark jets (``c jets'') are identified by the presence of corresponding hadrons
containing a b or c quark
among the ancestors of the jet constituents.
In the case where two jets contain the decay products of the same b hadron, the jet with the higher \pt is selected as the b jet.
When a b hadron is successfully matched, the c quarks are not considered.

The \ttjj sample is composed of four components, distinguished by the
flavour of the two jets in addition to the two b jets required from the
top quark decays.  The four components are the \ttbb final state with two
b jets, the \ttb final state with one b jet and one lighter-flavour jet,
the \ttcc final state with two c jets, and the \ttll final state with two
light-flavour jets (from a gluon or u, d, or s quark) or one light-flavour
jet and one c jet.
The \ttb final state is mainly from the merging of two b jets
or the loss of one of the b jets caused by the acceptance requirements.
Efficiency corrections to the measurement for the visible phase space are mainly from detector effects.
The results for the visible phase space are compared with those from MC simulations.

The goal of the full phase space result is to provide a comparison to
theoretical calculations, which are generally performed at the parton level.
To obtain a full phase space MC sample, the jet reconstruction is
performed on the partons (gluons, as well as quarks lighter than top) before
hadronization, as well as $\tau$ leptons that decay hadronically.  As the
full hadronization and decay chain is known, only $\tau$ leptons that
decay hadronically and partons that lead to hadrons are included.
The jet reconstruction algorithm is the same as for the visible phase space.
Following the jet reconstruction, b jets are
identified with a $\DR < 0.5$
requirement between the b quarks and parton-level jets, where $\Delta \phi$
and $\Delta \eta$ are the azimuthal angle and pseudorapidity
differences, respectively, between the directions of the b quark and the
parton-level jet.
For comparison with theoretical predictions~\cite{ttbbQCD},
results are quoted for two different jet \pt thresholds of
$\pt > 20$ and $>$ 40\GeVc on the jets not arising from top quark decays.
To clarify the phase space definition, the objects on which the selections are applied
are listed in Table~\ref{tab:ps}.

\begin{table*}
\centering
\topcaption{
The objects used to define the visible and full phase space are listed.
Details of the parton- and particle-level definitions are described in the text.
The symbol $\PQt$ denotes a top quark.
}
\label{tab:ps}
\begin{tabular}{l|cc}

Phase Space (PS)               &  parton level                                                  &  particle level                        \\  \hline
Visible PS                     &       ---                                                        &  4 (b) jets and 2 leptons ($\Pe, \mu$)   \\
Full PS                        &  $\PQt$, $\PAQt$  and 2 (b) jets (not from $\PQt$ or $\PAQt$)    &       ---                                \\

\end{tabular}
\end{table*}

\section{Event selection and background estimation}
The events are recorded using dilepton triggers with asymmetric thresholds of 8 and 17\GeVc on the
transverse momentum of the leptons.
Jets are reconstructed using the same algorithm as in the simulations.
The leptons and all charged hadrons that are associated with jets are required to originate from the primary vertex,
defined as the vertex with the highest $\sum{\pt^2}$ of its associated tracks.
Muon candidates are reconstructed by combining information from the silicon tracker and the muon system~\cite{paper:muon}.
Muon candidates are further required to have a minimum number of hits in the silicon tracker
and to have a high-quality global fit including a minimum number of hits in the muon detector.
Electron candidates are reconstructed by combining a track with energy deposits in the
ECAL, taking into account bremsstrahlung photons.
Requirements on electron identification variables based on shower shape and track-cluster matching are applied to the reconstructed candidates~\cite{PAS_PFT-10-003,PAS_EGM-10-001}.
Muons and electrons must have $\pt > 20$\GeVc and $\abs{\eta}< 2.4$.

To reduce the background contributions of muons or electrons from semileptonic heavy-flavour decays, relative isolation criteria are applied.
The relative isolation parameter, $I_\text{rel}$, is defined as the ratio of the sum of the transverse momenta of all objects in a cone of $\DR<0.3$
around the lepton \pt direction to the lepton \pt. The objects considered are the
charged hadrons associated with the primary vertex as well as
the neutral hadrons and photons, whose energies are corrected for the energy from pileup.
Thus,
\begin{equation}
I_\text{rel}=\frac{\sum{\pt^\text{charged hadron}} + \sum{\pt^\text{neutral hadron}} + \sum{\pt^\text{photon}}}{\pt^\text{lepton}}.
\end{equation}
Leptons are required to have $I_\text{rel}< 0.15$.
The efficiencies for the above lepton identification requirements are
measured using \Z boson candidates in data and are found to be consistent with the values from the simulation.
The residual differences are applied as a correction to the simulation.

The event selection requires the presence of two isolated opposite-sign leptons of invariant mass $M_{\ell\ell} > 12$\GeVcc.
Lepton pairs of the same flavour (\EE, \MM) are rejected if their invariant mass is within 15\GeVcc of the \Z boson mass.
The missing transverse energy (\ETmiss) is defined as the magnitude of the vectorial sum of the transverse momenta of all reconstructed particles in the event~\cite{paper:met}.
In the same-flavour channels, remaining backgrounds from $Z/\gamma^{*}$+jets processes are suppressed by demanding $\ETmiss> 30\GeV$.
For the \emu channel, no $\ETmiss$ requirement is applied.

Four or more reconstructed jets are required with $\abs{\eta}< 2.5$ and $\pt > 30$\GeVc, of which at least two jets must be identified as b jets,
using a combined secondary vertex (CSV) algorithm, which combines secondary vertex information with lifetime information of single tracks to produce a b-tagging discriminator~\cite{paper:btagid}.
A tight b-tagging requirement on this discriminator is applied, which has an efficiency of about 45\% for b jets and a misidentification probability of 0.1\% for light-flavour jets.

Differences in the b-tagging efficiencies between data and simulation~\cite{paper:btagid} are accounted for by reweighting the shape of the CSV b-tagging discriminator
distribution in the simulation to match that in the data. Data/MC scale factors for this \pt- and $\eta$-dependent correction are derived separately for light- and heavy-flavour jets.
The scale factor for c jets is not measured, owing to the limited amount of data, and is set to unity.
Light-flavour scale factors are determined from a control sample enriched in events with a \Z boson
and exactly two jets. Heavy-flavour scale factors are derived from a \ttbar enriched sample with exactly two jets,
excluding $\Z \to \ell\ell$ events.

The background contributions arising from $\cPZ/\gamma^{*}$+jets events is estimated in data using the number of events
having a dilepton invariant mass of $76< M_{\ell\ell}<106$\GeVcc,
scaled by the ratio of events that fail and pass this selection in the Drell--Yan simulation~\cite{top10001,CMStopPublication2}. The multijet and diboson background contributions are negligible after the full event selection.

\section{Measurement}
\begin{figure}[hbtp]
\centering
    \includegraphics[width=0.45\textwidth]{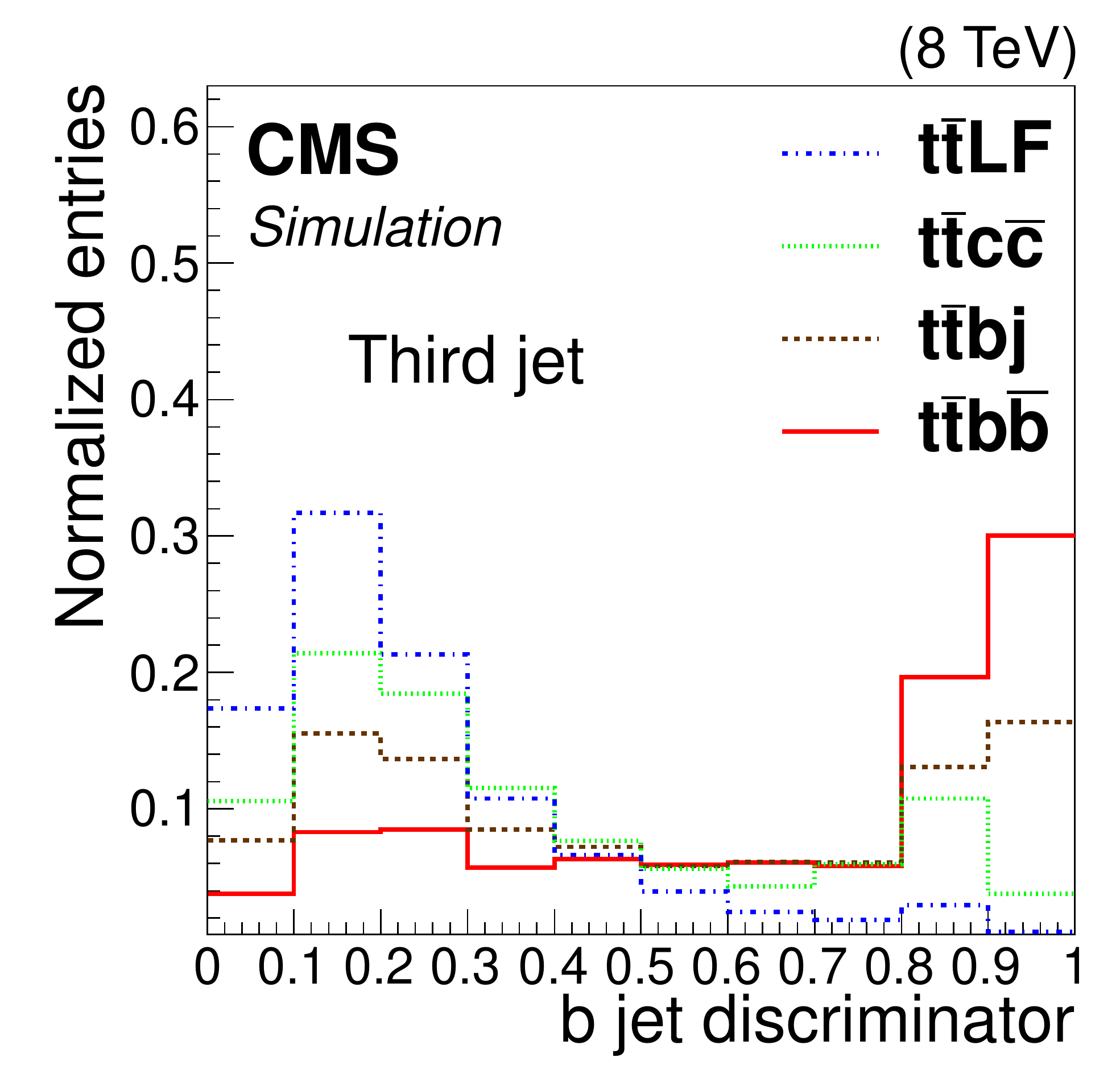}
    \includegraphics[width=0.45\textwidth]{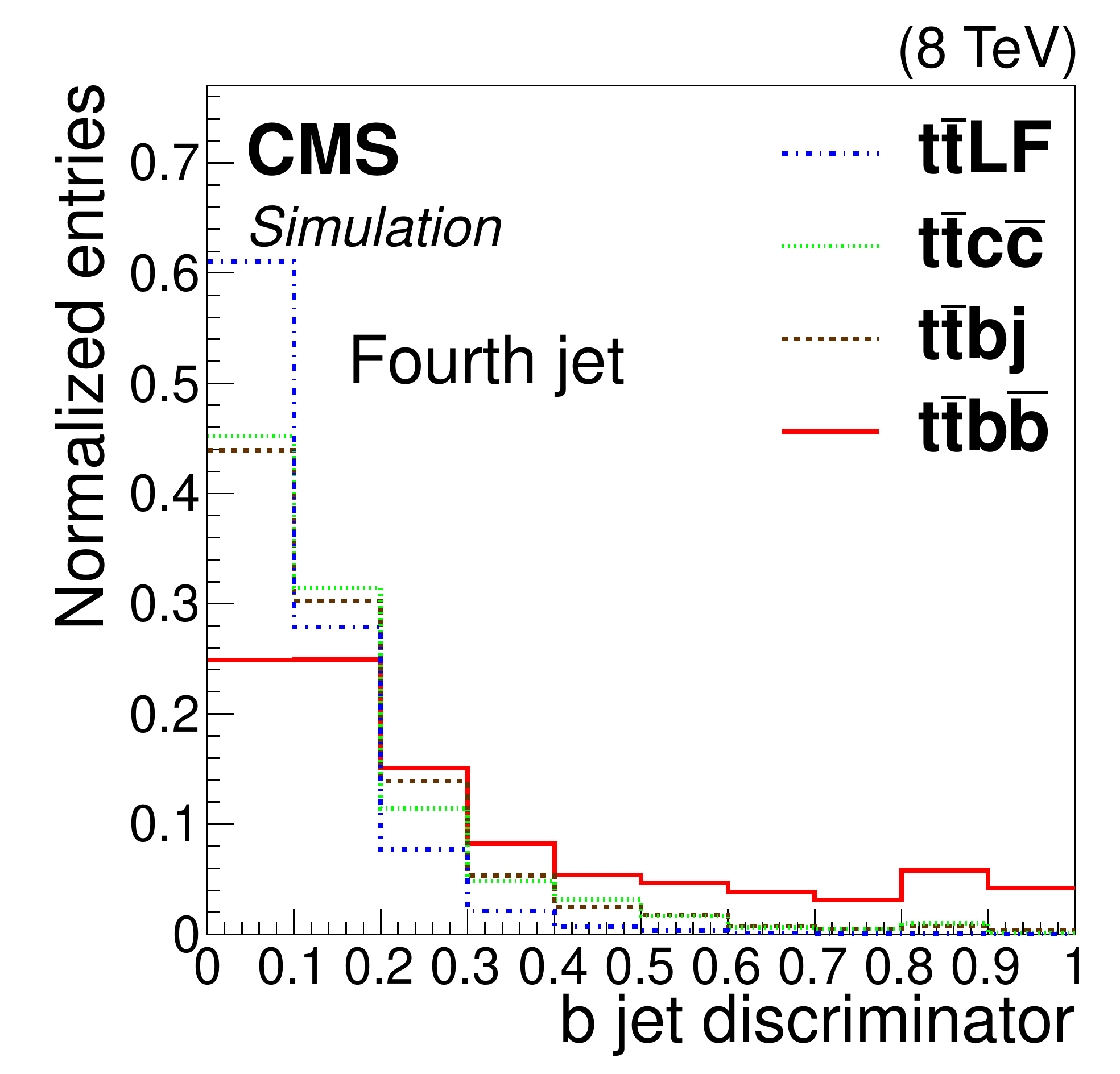}\\
    \caption{
Normalized distributions of the b jet discriminator for the third
(\cmsLeft) and fourth (\cmsRight) jets in an event, sorted in decreasing order of
b-tagging discriminator value, after the full event selection.  The
histograms are obtained from MC simulation and are separated according to
jet flavour.
}
    \label{fig:temp}

\end{figure}

After the full event selection, the three dilepton categories $\Pe\Pe$, $\mu\mu$, and $\Pe\mu$ are combined, and
the ratio of the number of \ttbb events to \ttjj events is obtained from the data by
fitting the CSV b-tagging discriminator distributions.
The distributions of the discriminator from simulation for the third and fourth jets in decreasing order of
the b-tagging discriminator, \ie for the two additional jets not identified as coming from the top quark decays,
are shown in Fig.~\ref{fig:temp}.
The third and fourth jets from \ttjj events tend to be light-flavour jets, while these are heavy-flavour jets for \ttbb events.
These two distributions are used to separate \ttbb from other processes.

\begin{figure}[hbtp]
\centering
    \includegraphics[width=0.45\textwidth]{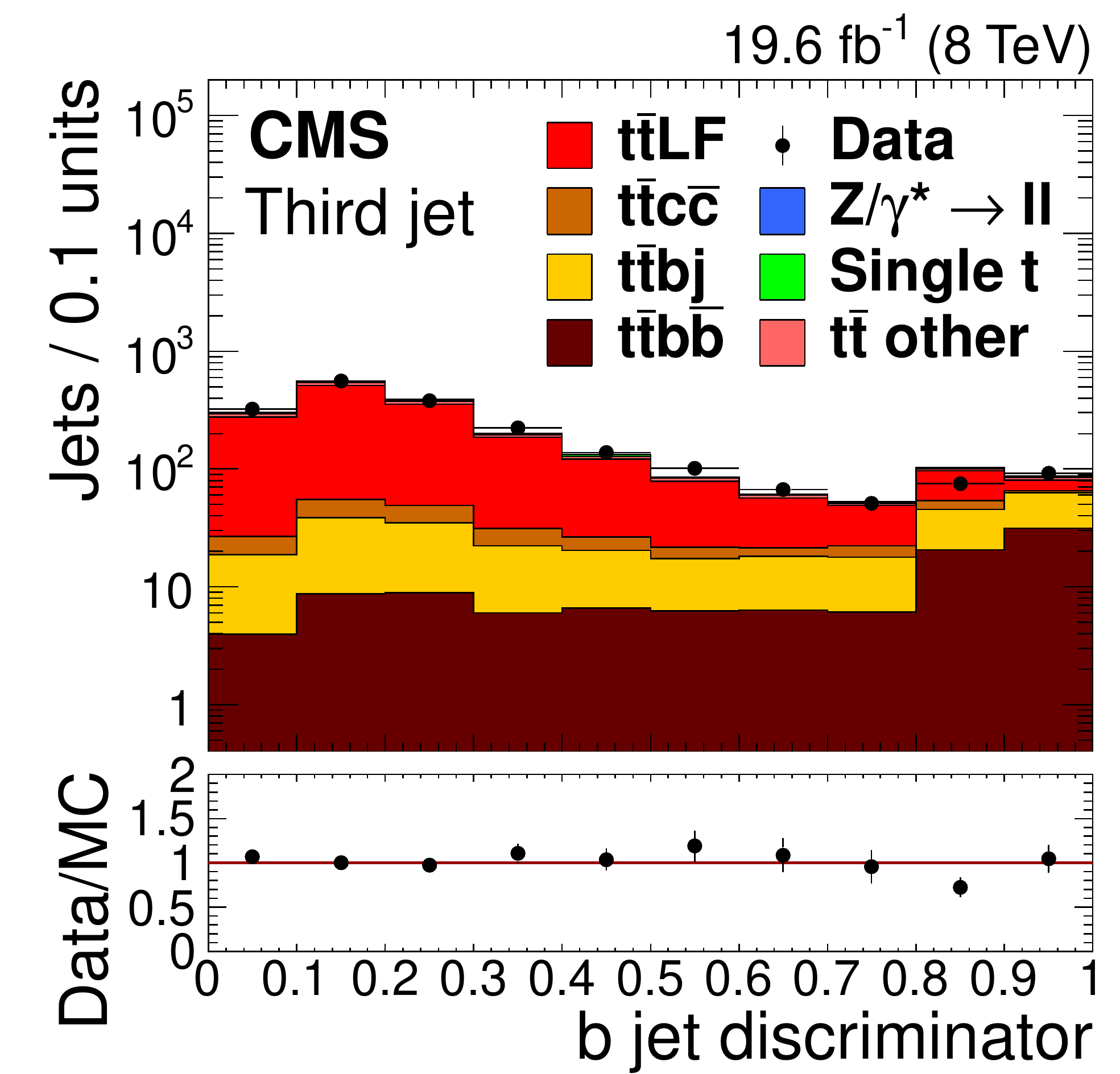}
    \includegraphics[width=0.45\textwidth]{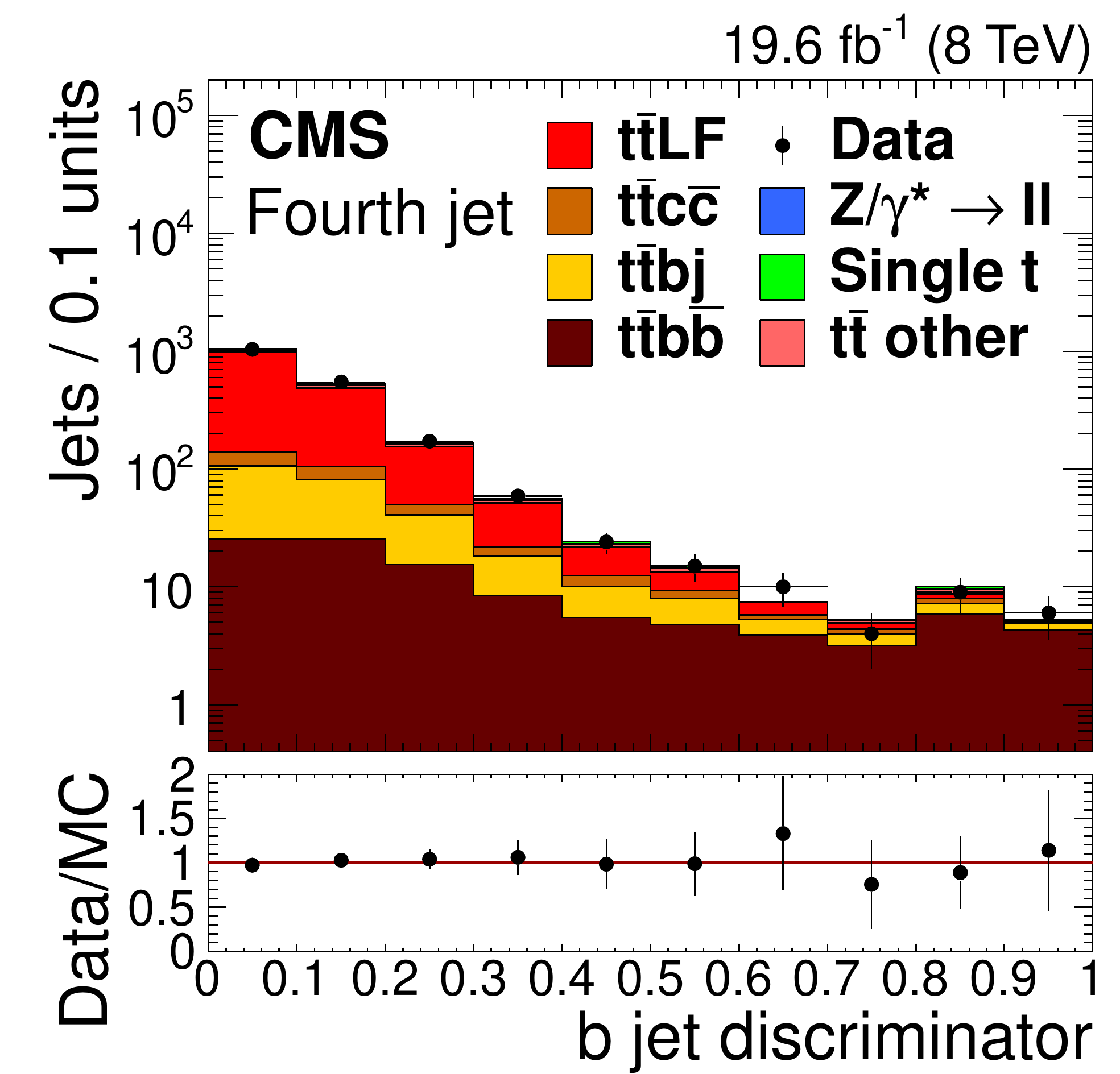}
    \caption{
Distributions of b jet discriminator for the third (\cmsLeft) and fourth (\cmsRight) jets in events
in decreasing order of b-tagging discriminator value, after the full event selection.
Points are from data and stacked histograms from MC simulation using results from the fit to data.
The ratio of the number of data events to the total number of MC events after the fit is shown in the lower panels.
}
    \label{fig:fitresult}

\end{figure}

Figure~\ref{fig:fitresult} shows the b-tagging discriminator distributions of the third
and fourth jets in the events from data and simulation, where the
simulation histograms have been scaled to the fit result.  The fit is
performed on both distributions simultaneously, and contains two free
parameters, an overall normalization and the ratio of the number of $\ttbb$ events to
$\ttjj$ events.  The $\ttcc$ and $\ttll$ contributions are combined, and
the ratio of the $\ttbb$ to $\ttb$ contributions is constrained using the predictions from
the MC simulation.  Additionally, the background contributions from
single top production and from \ttbar events that fail the visible phase space
requirements (labelled ``tt other'') are scaled by the normalization
parameter.  The contribution from $Z/\gamma^*$+jets is fixed from data,
as described above.  Nuisance parameters are used to account for the
uncertainties in the background contributions.

\begin{figure}[hbtp]
\centering
    \includegraphics[width=0.45\textwidth]{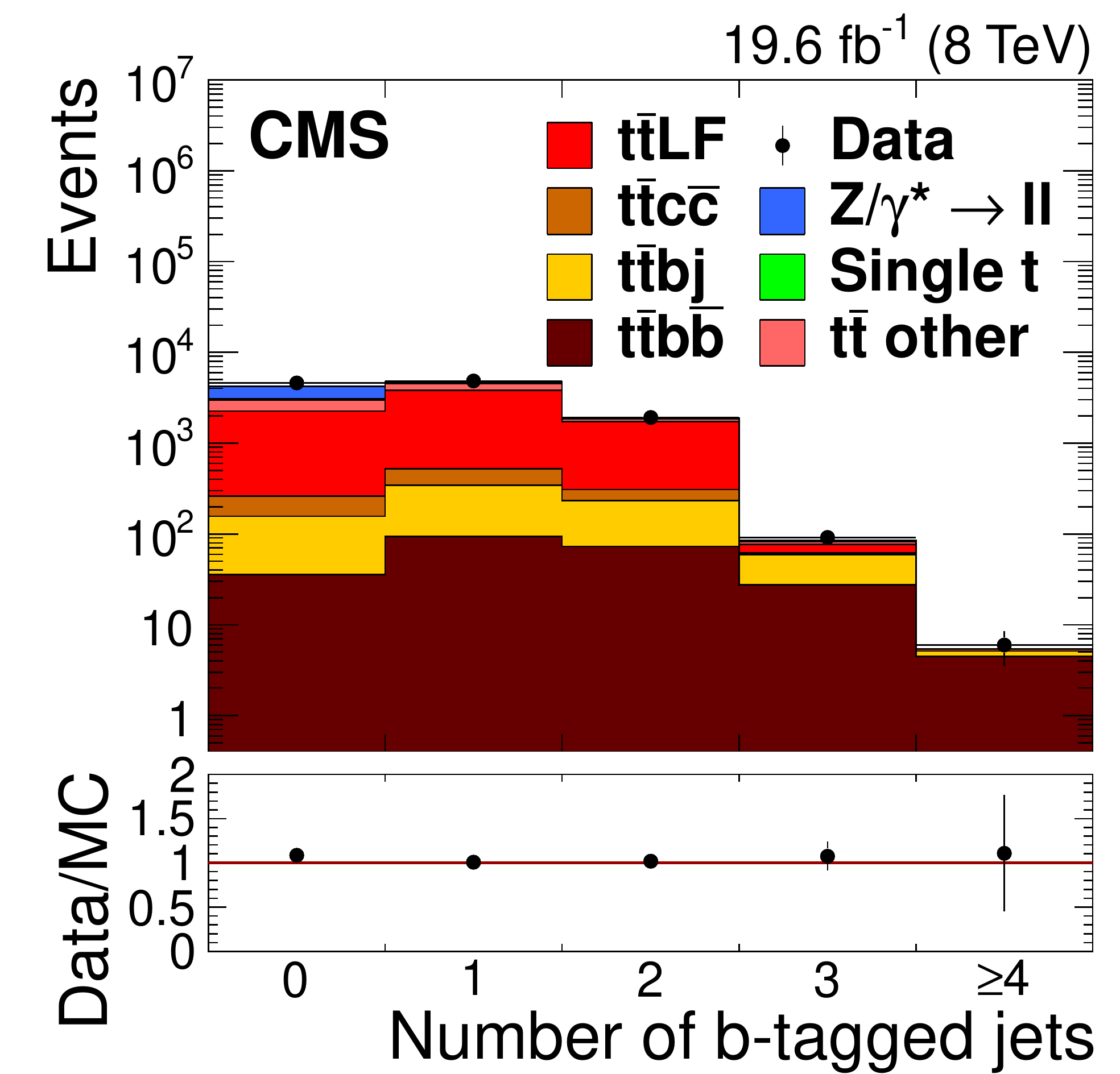}
    \caption{
Distribution of b jet multiplicity after the four-jet requirement but without the b-tagging requirement.
Points are from data and stacked histograms from MC simulation using results from the fit to data.
The ratio of the number of data events to the total number of MC events after the fit is shown in the lower panel.
}
    \label{fig:btag}

\end{figure}

The b-tagged jet multiplicity distribution in Fig.~\ref{fig:btag} shows
the comparison between data and the MC simulation, scaled by the fit results to the data.
The results, which include the requirement of four jets but not the b-tagging
requirement, indicate that the fit is a good match to the data, as made
clear in the lower panel showing the data/MC ratio.

Table~\ref{table:selfinal} gives the predicted number of events for each physics process
and for each dilepton category
after fitting to the data, as well as a comparison of the total number of
events expected from the simulation and observed in data.
Since the full event selection requires at least two b-tagged jets, which is usually satisfied by
\ttbar events, only 3\% of the events are from non-\ttbar processes.
The expected contribution from the \ttH process is 12 events.
This contribution is not subtracted from the data.

\begin{table}[h]
\topcaption{
The number of events for each physics process and for each dilepton category after
fitting to the data, their total, and the observed total number of events.  The
results are after the final event selection.  The $\cPZ/\gamma^* \to \ell\ell$
uncertainty is from data, while all other uncertainties include
only the statistical uncertainties in the MC samples.
}
\centering
\begin{tabular}{lcccr@{~$\pm$~}l}
Final state                      & \EE          &   \MM          &  \emu      &  \multicolumn{2}{c}{All}  \\ \hline
\ttbb                            & 18           &   26           &  61          & 105 & 2  \\  \hline
\ttb                             & 35           &   48           &  109         & 191 & 3  \\
\ttcc                            & 13           &   19           &  45          & 78 & 2    \\
\ttll                            & 249          &   347          &  840         & 1438 & 9  \\
\ttbar~ others                   & 21           &   25           &  64          & 109 & 3   \\
Single top                       & 7.4          &   11           &  24          & 43 & 5   \\
$Z/\gamma* \rightarrow \ell\ell  $ & 5.7          &   5.4          &  3.1         & 14 & 7   \\
\hline
Total                            & 350          & 483            & 1149         & 1983 & 13   \\
\hline
Data                       & 367          & 506            & 1145         & \multicolumn{2}{c}{2018}            \\
\end{tabular}

\label{table:selfinal}
\end{table}

The ratio of the number of \ttbb to \ttjj events at the reconstruction level obtained from the fit is corrected for the ratio of efficiencies.
The event selection efficiencies, defined as the number of  \ttbb and \ttjj events after the full event selection
divided by the number of events
in the corresponding visible phase space are 18.7\% and 7.2\%, respectively.
The \ttbb and \ttjj cross sections in the visible phase space are measured using $\sigma_\text{visible} =  N / ( \epsilon  \mathcal{L} ) $,
where $\mathcal{L}$ is the integrated luminosity, $N$ is the number of observed events,
and $\epsilon$ is the efficiency for each process.
However, the NLO theoretical calculation is based on parton-level jets being clustered with partons before hadronization in the full phase space.
For the purpose of comparing with the theoretical prediction, the cross sections in the full phase space are extrapolated from the cross sections
in the visible phase space
using $\sigma_\text{full} = \sigma_\text{visible} / \mathcal{A}$,
where $\mathcal{A}$ is the acceptance.
The acceptances for extending \ttbb and \ttjj
to the full phase space based on the \MADGRAPH simulation are 2.6\% and 2.4\%, respectively,
including the \ttbar to dilepton branching fraction, calculated using
the leptonic branching fraction of the W boson~\cite{PDG2014}.
The acceptance is defined as the number of events in the corresponding visible phase space divided by the number of events in the full phase space.
\section{Estimation of systematic uncertainties}
The systematic uncertainties are determined separately for the \ttbb and \ttjj cross sections and their ratio.
In the ratio, many systematic effects cancel,
specifically normalization uncertainties such as the ones related to the measurement of the integrated luminosity and
the lepton identification including trigger efficiencies, since they are common to both processes.
The various systematic uncertainties in the measured values are
shown in Table~\ref{tab:sys} for the visible phase space and a jet \pt threshold of 20\GeVc, including the luminosity
uncertainty~\cite{CMS-PAS-LUM-13-001} and lepton identification~\cite{CMStopdilepton8TeV},
which only affect the absolute cross section measurements.
The systematic uncertainty in the lepton identification is assessed
using the scale factor obtained from \Z boson candidates and also taking into account the different phase space between
\Z boson and \ttbar events.

The systematic uncertainties associated with the b-tagging discriminator scale factors for b jets and light-flavour jets
are studied separately, varying their values within their uncertainties.
The b-flavour scale factors are obtained using \ttbar enriched events, and their dominant uncertainty comes from the contamination when
one of the b jets is not reconstructed~\cite{cms:ttH} (indicated as ``b quark flavour'' in Table~\ref{tab:sys}).
The c jet scale factor is assumed to be unity with
an uncertainty twice as large as the b-tagging scale factor~\cite{paper:btagid} (indicated as ``c quark flavour'' in Table~\ref{tab:sys}).
The light-flavour jet scale factors are determined from \Z boson enriched events.
Their uncertainty arises because the contribution from the \Z + \bbbar process in this control sample is not well modelled (indicated as ``light flavour'' in Table~\ref{tab:sys}).
The b-tagging discriminator can be affected by the jet energy scale (JES) variations.
The systematic uncertainty in the jet energy scale~\cite{paper:jet} is obtained by varying
the jet energy scale factor by one standard deviation for each quark flavour.
The uncertainty in the jet energy resolution (JER) is assessed by
smearing the simulated jet energy resolution by 10\%
on average, taking into account the $\eta$ dependence~\cite{paper:jet}.

The uncertainty arising from
constraining the ratio of the $\ttb$ to $\ttbb$ contributions in the fit
to match the MC prediction
is evaluated by comparing the result with and without the constraint.
The number of pileup interactions in data is estimated from
the measured bunch-to-bunch instantaneous luminosity and the total inelastic cross section.
The systematic uncertainty in the number of pileup events is estimated by conservatively
varying this cross section by 5\%
to cover all the uncertainties in the modelling of the pileup physics.
The contributions from Drell-Yan and single top quark processes are small, and the shapes of the distributions from
these backgrounds are similar to those of the \ttll component.
Therefore, these backgrounds do not affect the measurement significantly.
For the efficiency of \ttjj events, the uncertainty owing to the heavy-flavour fraction is estimated by varying the contribution by 50\%.
An uncertainty to account for the variation of the \ttcc fraction in the fit is also assigned by varying the contribution by 50\%.
This variation is chosen because the theoretical uncertainty in the \ttjj cross section is less than 50\%,
and the fitted \ttcc fraction remains within 50\% of the input value when fitting with the \ttcc contribution as a free parameter.

The dependence of the correction factor for the particle level on the
assumptions made in the MC simulation is another source of systematic
uncertainty: the generators \MADGRAPH and \POWHEG are compared and the difference in the efficiency ratio is taken as the systematic uncertainty.
The uncertainties from the
factorization/renormalization scales and the matching scale that
separates jets from ME and from parton showers in \MADGRAPH are estimated
by varying the scales a factor of two up and down with respect to their
reference values.
The uncertainties in the PDFs are accounted for by following the
PDF4LHC prescription~\cite{bib:PDF4LHC}.

The total systematic uncertainty in the cross section
ratio is 22\%, with the dominant contributions from the b-tagging efficiency and the
misidentification of light-flavoured partons, followed by the
renormalization/factorization and matching scale systematic
uncertainties.

The uncertainty in \sttjj is significantly smaller than that in \sttbb since  the measurement of the latter requires the identification of multiple b jets.
The uncertainty in \sttbb
is larger than that for the cross section ratio since uncertainties that are common between \ttbb and \ttjj,
such as the jet energy scale uncertainty, partially or completely cancel in the ratio.

The systematic uncertainties in the measurements with a \pt threshold of 40\GeVc are found to be very similar to
those with a 20\GeVc threshold.
The uncertainty from the factorization and renormalization scales for the higher-\pt threshold of 40\GeVc
cannot be accurately determined owing to the statistical uncertainties in the MC sample. Thus, the $\pt > 40$\GeVc
threshold measurements use the same scale ($\mu_F$ and $\mu_R$) systematic uncertainties
as those found for the $\pt > 20$\GeVc threshold results.

In extrapolating the measurements from the visible phase space to the full phase space,
the systematic uncertainty in the acceptance is included.
The effect of the MC modelling of the acceptance is estimated by comparing
the results between \MADGRAPH and \POWHEG.
This uncertainty equals 5\% for each of the cross section measurements and 2\% for the cross section ratio.

\begin{table}
\centering
\topcaption{Summary of the systematic uncertainties from various sources contributing to $\sttbb$, $\sttjj$, and the ratio  $\sttbb/\sttjj$
for a jet \pt threshold of $\pt > 20$\GeVc in the visible phase space.
 }
\label{tab:sys}
\begin{tabular}{l|c|c|c}
                        Source                          &   $\sttbb$\,(\%)      &  $\sttjj$\,(\%)   &  $\frac{\sttbb}{\sttjj}$\,(\%)   \\
                        \hline
                        Pileup                          &    1.0           &    1.0     &   1.0                   \\
                        JES \& JER                      &    11            &    8.0     &   5.0                   \\
                        b tag (b quark flavour)          &    15            &    $<$0.1 &   15                    \\
                        b tag (c quark flavour)          &    4.0           &    $<$0.1 &   4.0                   \\
                        b tag (light flavour)            &    7.0           &    $<$0.1 &   7.0                   \\
                        Ratio of \ttbb and \ttb         &    9.0           &    $<$0.1 &   9.0                   \\
                        Bkgnd modelling             &    1.0           &    1.0     &   1.0                   \\
                        \ttcc fraction in the fit       &    4.2           &    0.2     &   4.0                   \\
                        Lepton identification            &    4.0           &    4.0     &   ---                   \\
                        \hline
                        MC generator                    &    3.0           &    3.0     &   3.0                   \\
                        Scale ($\mu_F$ and $\mu_R$)     &    8.0           &    3.0     &   6.0                   \\
                        PS matching                     &    12            &    5.0     &   3.0                   \\
                        PDF                             &    4.0           &    4.0     &   $<$0.1                   \\
                        Eff. (\ttcc fraction)     &    ---            &    1.6     &   1.6                   \\
                        \hline
                        Luminosity                      &    2.6           &    2.6     &    ---                   \\
                        \hline
                        Total uncertainty               &    28            &    12      &   22                    \\
\end{tabular}

\end{table}

\section{Results}
After correcting for the efficiency ratio and taking into account the systematic uncertainties,
the cross section ratio \sttbb/\sttjj\ is measured in the visible phase space
from a fit to the measured CSV b-tagging discriminator distributions shown in Fig.~\ref{fig:fitresult}.
The measured cross section ratio in the visible phase space for events with particle-level jets and a minimum jet \pt of 20\GeVc is
\begin{equation}\label{eq:20GeV}
\sttbb/\sttjj=0.022 \pm 0.003\stat \pm 0.005\syst.
\end{equation}
This result is for the visible phase space, defined as events having two leptons with
$\pt >20\GeVc$ and $\abs{\eta}<2.4$, plus four jets, including two b jets with $\pt>20\GeVc$ and
$\abs{\eta}<2.5$.  The predicted value from both \MADGRAPH and \POWHEG is
found to be $0.016 \pm 0.002$,
where the MC uncertainty is
the sum in quadrature of the statistical uncertainty and the systematic uncertainties from
the factorization/renormalization and the matching scales.
The measured cross sections are presented in Table~\ref{tab:ttbbttjjX}.
When the \ttH contribution is subtracted from the data, the ratio is reduced by only 4\%, much less than the overall uncertainty.
Therefore, compared to the uncertainties, the contribution from \ttH can be considered negligible.
The measured full phase space ratio with a minimum \pt of 20\GeVc for parton-level jets is consistent within the uncertainties
with the result in the visible phase space.

A NLO theoretical QCD calculation is available for parton-level jets with a $\pt > 40$\GeVc threshold~\cite{ttbbQCD}.
The NLO cross section values for $\sttbb$, $\sttjj$,
and the ratio \sttbb/\sttjj\ are given in Table~\ref{tab:ttbbttjjX}.
To compare with this theoretical prediction, the analysis is repeated for a jet threshold of $\pt > 40$\GeVc.
Correspondingly with a higher jet \pt threshold in the event selection,
24 $\ttbb$ events and 478 $\ttjj$ events remain after the full event selection,
with the acceptance (including the event selection efficiency) of 0.34\% and 0.15\%, respectively.
The measured cross section ratio in the full phase space with the $\pt > 40$\GeVc threshold is
\begin{equation}
\sttbb/\sttjj = 0.022 \pm 0.004\stat \pm 0.005\syst.
\end{equation}
The cross sections in the full phase space for this \pt threshold are summarized in Table~\ref{tab:ttbbttjjX}.
The measured cross section ratio is higher, but compatible within 1.6 standard deviations with the prediction from the NLO calculation
of $0.011 \pm 0.003$.

\begin{table*}
\centering
\topcaption{
The measured cross sections \sttbb\ and \sttjj\ and their ratio are given
for the visible phase space (PS) defined as two leptons with
$\pt>20\GeVc$ and $\abs{\eta}<2.4$ plus four jets, including two b jets with $\pt>20\GeVc$ and
$\abs{\eta}<2.5$, and the full phase space, corrected for acceptance and
branching fractions.  The full phase space results are given for jet
thresholds of $\pt>20$ and 40\GeVc.
The uncertainties shown are statistical and systematic, respectively.
The predictions of a NLO theoretical calculation for the full phase space and $\pt > 40$\GeVc are also given~\cite{ttbbQCD}.
}
\label{tab:ttbbttjjX}
\begin{tabular}{l|ccc}
Phase Space (PS)               &   $\sttbb$\,[pb]                         &  $\sttjj$\,[pb]                               &  \sttbb/\sttjj \\ \hline
Visible PS (particle)          &                                         &                                              &        \\
Jet $\pt > 20$\GeVc           &   0.029 $\pm$ 0.003 $\pm$ 0.008         &  1.28 $\pm$ 0.03 $\pm$ 0.15 &  0.022 $\pm$ 0.003 $\pm$ 0.005 \\ \hline
Full PS (parton)               &                                         &                                 &                      \\
Jet $\pt > 20$\GeVc           &   1.11 $\pm$ 0.11 $\pm$ 0.31     &  52.1 $\pm$ 1.0 $\pm$ 6.8     &  0.021 $\pm$ 0.003 $\pm$ 0.005\\
Jet $\pt > 40$\GeVc           &   0.36 $\pm$ 0.08 $\pm$ 0.10     &  16.1 $\pm$ 0.7 $\pm$ 2.1     &  0.022 $\pm$ 0.004 $\pm$ 0.005\\[0.5ex]
NLO calculation                &                                         &                                 &                       \\
Jet $\pt > 40$\GeVc           &   0.23 $\pm$ 0.05                       &  21.0 $\pm$ 2.9                 &  0.011 $\pm$ 0.003   \\
\end{tabular}

\end{table*}

\section{Summary}
A measurement of the cross section ratio \sttbb/\sttjj has been presented by the CMS experiment,
using a data sample of pp collisions at a centre-of-mass energy of 8\TeV, corresponding to an integrated luminosity of 19.6\fbinv.
The individual cross sections \sttjj and \sttbb have also been determined.
The cross section ratio was measured in a visible phase space region using the dilepton decay mode of \ttbar events and corrected to the particle level,
corresponding to the detector acceptance.
The measured cross section ratio in the visible phase space is $\sttbb/\sttjj = 0.022\pm0.003\stat\pm0.005\syst$
with a minimum \pt for the particle-level jets of 20\GeVc.
The cross section ratio has also been measured in the full phase space with minimum parton-jet \pt thresholds of $\pt >20$ and $>$40\GeVc
in order to compare with a NLO QCD calculation of the cross section ratio.
The measurement is compatible within 1.6 standard deviations with the theoretical prediction.
These are the first measurements of the cross sections $\sttbb$ and $\sttjj$, and their ratio.
The result will provide important information about the main background in the search for \ttH
and as a figure of merit for testing the validity of NLO QCD calculations.

\begin{acknowledgments}
We congratulate our colleagues in the CERN accelerator departments for the excellent performance of the LHC and thank the technical and administrative staffs at CERN and at other CMS institutes for their contributions to the success of the CMS effort. In addition, we gratefully acknowledge the computing centres and personnel of the Worldwide LHC Computing Grid for delivering so effectively the computing infrastructure essential to our analyses. Finally, we acknowledge the enduring support for the construction and operation of the LHC and the CMS detector provided by the following funding agencies: BMWFW and FWF (Austria); FNRS and FWO (Belgium); CNPq, CAPES, FAPERJ, and FAPESP (Brazil); MES (Bulgaria); CERN; CAS, MoST, and NSFC (China); COLCIENCIAS (Colombia); MSES and CSF (Croatia); RPF (Cyprus); MoER, ERC IUT and ERDF (Estonia); Academy of Finland, MEC, and HIP (Finland); CEA and CNRS/IN2P3 (France); BMBF, DFG, and HGF (Germany); GSRT (Greece); OTKA and NIH (Hungary); DAE and DST (India); IPM (Iran); SFI (Ireland); INFN (Italy); MSIP and NRF (Republic of Korea); LAS (Lithuania); MOE and UM (Malaysia); CINVESTAV, CONACYT, SEP, and UASLP-FAI (Mexico); MBIE (New Zealand); PAEC (Pakistan); MSHE and NSC (Poland); FCT (Portugal); JINR (Dubna); MON, RosAtom, RAS and RFBR (Russia); MESTD (Serbia); SEIDI and CPAN (Spain); Swiss Funding Agencies (Switzerland); MST (Taipei); ThEPCenter, IPST, STAR and NSTDA (Thailand); TUBITAK and TAEK (Turkey); NASU and SFFR (Ukraine); STFC (United Kingdom); DOE and NSF (USA).

Individuals have received support from the Marie-Curie programme and the European Research Council and EPLANET (European Union); the Leventis Foundation; the A. P. Sloan Foundation; the Alexander von Humboldt Foundation; the Belgian Federal Science Policy Office; the Fonds pour la Formation \`a la Recherche dans l'Industrie et dans l'Agriculture (FRIA-Belgium); the Agentschap voor Innovatie door Wetenschap en Technologie (IWT-Belgium); the Ministry of Education, Youth and Sports (MEYS) of the Czech Republic; the Council of Science and Industrial Research, India; the HOMING PLUS programme of Foundation for Polish Science, cofinanced from European Union, Regional Development Fund; the Compagnia di San Paolo (Torino); the Consorzio per la Fisica (Trieste); MIUR project 20108T4XTM (Italy); the Thalis and Aristeia programmes cofinanced by EU-ESF and the Greek NSRF; and the National Priorities Research Program by Qatar National Research Fund.
\end{acknowledgments}

\bibliography{auto_generated}   

\cleardoublepage \appendix\section{The CMS Collaboration \label{app:collab}}\begin{sloppypar}\hyphenpenalty=5000\widowpenalty=500\clubpenalty=5000\textbf{Yerevan Physics Institute,  Yerevan,  Armenia}\\*[0pt]
V.~Khachatryan, A.M.~Sirunyan, A.~Tumasyan
\vskip\cmsinstskip
\textbf{Institut f\"{u}r Hochenergiephysik der OeAW,  Wien,  Austria}\\*[0pt]
W.~Adam, T.~Bergauer, M.~Dragicevic, J.~Er\"{o}, C.~Fabjan\cmsAuthorMark{1}, M.~Friedl, R.~Fr\"{u}hwirth\cmsAuthorMark{1}, V.M.~Ghete, C.~Hartl, N.~H\"{o}rmann, J.~Hrubec, M.~Jeitler\cmsAuthorMark{1}, W.~Kiesenhofer, V.~Kn\"{u}nz, M.~Krammer\cmsAuthorMark{1}, I.~Kr\"{a}tschmer, D.~Liko, I.~Mikulec, D.~Rabady\cmsAuthorMark{2}, B.~Rahbaran, H.~Rohringer, R.~Sch\"{o}fbeck, J.~Strauss, A.~Taurok, W.~Treberer-Treberspurg, W.~Waltenberger, C.-E.~Wulz\cmsAuthorMark{1}
% ...
\vskip\cmsinstskip
% ...
1:~~Also at Vienna University of Technology, Vienna, Austria\\
2:~~Also at CERN, European Organization for Nuclear Research, Geneva, Switzerland\\
% ...

\end{sloppypar}
\end{document}